\documentclass[pra]{revtex4}
\usepackage{amssymb}
\usepackage{amsmath}
\usepackage{graphicx}

\begin{document}

\title{Suppression of the quantum collapse in binary bosonic gases}
\author{Hidetsugu Sakaguchi$^{1}$ and Boris A. Malomed$^{2}$}
\affiliation{$^{1}$Department of Applied Science for Electronics and Materials,
Interdisciplinary Graduate School of Engineering Sciences, Kyushu
University, Kasuga, Fukuoka 816-8580, Japan\\
$^{2}$Department of Physical Electronics, School of Electrical Engineering,
Faculty of Engineering, Tel Aviv University, Tel Aviv 69978, Israel}

\begin{abstract}
Attraction of the quantum particle to the center in the 3D space with
potential $-V_{0}r^{-2}$ gives rise to the \textit{quantum collapse}, i.e.,
nonexistence of the ground state (GS) when the attraction strength exceeds a
critical value [$\left( V_{0}^{(\mathrm{cr})}\right) _{1}=1/8$, in the
present notation]. Recently, we have demonstrated that the quantum collapse
is suppressed, and the GS is restored, if repulsive interactions between
particles in the quantum gas are taken into account, in the mean-field
approximation. This setting can be realized in a gas of dipolar molecules
attracted to the central charge, with dipole-dipole interactions taken into
regard too. Here we analyze this problem for a binary gas. GSs supported by
the repulsive interactions are constructed in a numerical form, as well as
by means of analytical approximations for both miscible and immiscible
binary systems. In particular, the Thomas-Fermi (TF) approximation is
relevant if $V_{0}$ is large enough. It is found that the GS of the miscible
binary gas, both balanced and imbalanced, features a \emph{weak phase
transition} at another critical value, $\left( V_{0}^{(\mathrm{cr})}\right)
_{2}=1/2\equiv 4\left( V_{0}^{(\mathrm{cr})}\right) _{1}$. The transition is
characterized by an analyticity-breaking change in the structure of the wave
functions at small $r$. To illustrate the generic character of the present
phenomenology, we also consider the binary system with the attraction
between the species (rather than repulsion), in the case when the central
potential pulls a single component only.
\end{abstract}

\maketitle

\section{Introduction}

Attractive potential $U(r)=-V_{0}r^{-2}$ plays a critical role for the
effect of the quantum collapse (alias ``fall onto the center") in the
three-dimensional (3D) space \cite{LL}: If $V_{0}$ exceeds a critical value
[it is $V_{0}^{(\mathrm{cr})}=1/8$ in the notation adopted below, see Eqs. (%
\ref{GP})], the corresponding 3D Schr\"{o}dinger equation fails to produce
the ground state (GS) (loosely speaking, in this case the energy of the GS
collapses to $E\rightarrow -\infty $). This critical feature is explained by
the fact that the quantization breaks the scaling invariance of the
classical counterpart of the present quantum setting. The effect is also
known as the \textit{quantum anomaly}, alias ``dimensional transmutation"
\cite{anomaly}.

A solution of the problem of the missing GS was elaborated in terms of the
linear quantum field theory, which, by means of the renormalization
procedure, postulated the existence of the GS, with an \emph{arbitrary}
spatial size \cite{anomaly}. A different approach was recently proposed in
our work \cite{we1}, which considered, in terms of the mean-field
approximation, a bosonic gas (rather than the single particle) pulled to the
center by potential $-V_{0}r^{-2}$, and took into regard the
collision-induced repulsive nonlinearity in the corresponding
Gross-Pitaevskii equation (GPE) \cite{Pit}. The physical realization of the
setting is possible in ultra-cold gases of molecules carrying a permanent
electric dipole moment, $d$ (such as LiCs \cite{LiCs} and KRb \cite{KRb}),
which are attracted by electric charge $Q$ placed at the center (it may be
an ion held by an optical trapping potential \cite{ion}). In this case, the
attraction constant is%
\begin{equation}
V_{0}=|Q|d.  \label{V0}
\end{equation}

Alternatively, it is possible to use an atomic gas in a long-lived excited
state (such as a ``frozen" low-lying Rydberg state \cite%
{Rydberg,Rydberg-review}, in which the permanent dipole moment grows
with the principal quantum number, $n$, as $n^{2}$); to additionally
stabilize
the gas of excited atoms, one may pump it by a resonant laser field \cite%
{Rydberg-pump}. The analysis reported in Ref. \cite{we1} took into account
the dipole-dipole interaction in the gas too, also in the framework of the
mean-field approximation, which amounts to an effective renormalization
(increase) of the strength of the contact nonlinearity. As a result, it has
been demonstrated that the repulsive nonlinearity suppresses the quantum
collapse at all values of $V_{0}$, restoring the GS. An estimate with
relevant values of physical parameters demonstrates that the radius of the
resurrected GS may be a few $\mathrm{\mu }$m. In Ref. \cite{we2}, the
analysis was extended for the dipolar gas polarized by strong uniform dc
field, with the spatial symmetry reduced from spherical to cylindrical.

In Ref. \cite{we1}, the same problem was considered in the two-dimensional
(2D) geometry, where, in the framework of the linear Schr\"{o}dinger
equation, the attractive potential $-V_{0}r^{-2}$ leads to the collapse at
any finite value of $V_{0}$. In addition to the above-mentioned physical
implementations, the 2D gas subject to the action of this potential may be
composed of polarizable atoms without a permanent dielectric moment, while
an effective moment is induced in them by the electric field of a uniformly
charged wire set perpendicular to the system's plane \cite{Schmiedmayer1},
or with an effective magnetic moment induced by a current filament set in
the perpendicular direction \cite{we1}. However, the cubic repulsive
nonlinearity is too weak to suppress the quantum collapse in the 2D setting,
only the quintic term (if it is physically relevant) being strong enough for
that purpose \cite{we1}. Indeed, the experimental realization of the
above-mentioned quasi-2D system of polarizable atoms attracted to the
charged wire has demonstrated a collapse-like behavior \cite{Schmiedmayer2}.
Furthermore, if a gas is tightly confined near the 2D plane by a strong
trapping potential acting in the transverse direction, then, in the limit of
large density, the underlying cubic nonlinearity is effectively transformed
into that with power $7/3$ \cite{Delgado}, which is still weaker. On the
other hand, in the same limit the underlying quintic term will be
transformed into one with power $11/3$, which is sufficient for the
suppression of the quantum collapse and rebuilding of the GS.

The objective of the present work is to extend the analysis of the
suppression of the quantum collapse for the \emph{binary gas}, which
combines the intra-species repulsion with the interaction between two
species, the relative strength of the inter-species repulsion being $\gamma $%
, see Eqs. (\ref{GP}) below. It is well known that $\gamma $ can be adjusted
by means of the Feshbach resonance controlled by external fields \cite%
{Feshbach1,Feshbach2}. This, in turn, allows one to switch the binary system
between regimes of miscibility at $\gamma <1$ and immiscibility at $\gamma
>1 $ \cite{Mineev,misc}. The point of the miscibility-immiscibility
transition may be shifted from $\gamma =1$ to $\gamma >1$ by external
confinement \cite{Merhasin,confinement}, and by linear mixing between the
species \cite{Merhasin,linear-coupling}, including the mixing which
represents the effective spin-orbit coupling in binary condensates \cite{SO}%
. The immiscibility leads to the phase separation of binary condensates and
formation of domain walls in the 1D geometry \cite{DW}. In 2D and 3D
settings, domain walls often form circular or spherical shells, respectively
\cite{shells}.

In this \ work, we construct spherically symmetric GSs pulled to the center
by potential $-V_{0}r^{-2}$, which acts on both species, or a single one, in
the binary quantum gas. The GSs are stabilized against the collapse by the
repulsive nonlinearity (in the case when the single species carries the
dipolar moment pulled to the central charge, the inter-species interaction
must be attractive, corresponding to $\gamma <0$, rather than repulsive).
The GSs are constructed in miscible and immiscible settings alike. In the
case of the immiscibility, we could identify phase-separated structures in
the form of spherically symmetric shells, but not sectors (e.g., for equal
norms of the immiscible components, we have not found states in the form of
two hemispheres separated by a flat domain wall). In the miscible binary
gas, two critical values of $V_{0}$ are found. One is $\widetilde{V}_{0}^{(%
\mathrm{cr})}=1/2\equiv 4V_{0}^{(\mathrm{cr})}$ (recall $V_{0}^{(\mathrm{cr}%
)}=1/8$ is the value at which the GS breaks down in the linear
Schr\"{o}dinger equation) in the \emph{balanced} and
\emph{imbalanced} miscible gases alike, i.e., the gases with equal
or different norms of the two components. At $V_{0}=1/2$, the
coupled GPEs feature a breakup of the analyticity and a change in
the structure of the GS wave function at $r\rightarrow 0$
(nevertheless, the GS exists equally well at $V_{0}<1/2$ and
$V_{0}>1/2$). The imbalanced miscible gas (with $\gamma <1$) gives
rise to an additional critical
value,$~\mathcal{V}_{0}^{(\mathrm{cr})}=(1/2)\left( 1+\gamma \right)
/\left( 1-\gamma \right) $, which we consider only briefly.

The rest of the paper is organized as follows. The \ model is formulated in
Section II, which also includes analysis of the asymptotic form of the wave
functions. In section III, basic numerical results are reported for the
miscible and immiscible settings, along with additional approximate
analytical results, including the Thomas-Fermi (TF) approximation, which is
relevant in the case of large $V_{0}$. In Section IV, we consider the system
with the single component pulled to the center, and attractive, rather than
repulsive, inter-species interaction ($\gamma <0$). The paper is concluded
by Section V.

\section{Formulation of the model and analytical considerations}

The self-repulsive binary condensate pulled to the center by the attractive
potential is described by the coupled GPEs, written here in the scaled
form,following Ref. \cite{we1}:
\begin{eqnarray}
i\frac{\partial \phi _{1}}{\partial t} &=&-\frac{1}{2}\nabla ^{2}\phi
_{1}+(|\phi _{1}|^{2}+\gamma |\phi _{2}|^{2})\phi _{1}-\frac{V_{0}}{r^{2}}%
\phi _{1}~,  \notag \\
&&  \label{GP} \\
i\frac{\partial \phi _{2}}{\partial t} &=&-\frac{1}{2}\nabla ^{2}\phi
_{2}+(\gamma |\phi _{1}|^{2}+|\phi _{2}|^{2})\phi _{2}-\frac{V_{0}}{r^{2}}%
\phi _{2}~  \notag
\end{eqnarray}%
(in Ref. \cite{we1}, the strength of the attractive potential was defined as
$U_{0}\equiv 2V_{0}$), where $\gamma $ is the relative strength of the
inter-species repulsion, while coefficients of the self-repulsion are scaled
to be $1$. Spherically symmetric stationary states with chemical potential $%
\mu _{n}$, $n=1,2$, are looked for as
\begin{equation}
\phi _{n}\left( r,t\right) =r^{-1}\chi _{n}(r)\exp \left( -i\mu _{n}t\right)
,  \label{phi}
\end{equation}%
where $\chi _{n}(r)$ are real functions obeying the coupled equations,
\begin{eqnarray}
\mu _{1}\chi _{1} &=&-\frac{1}{2}\chi _{1}^{\prime \prime }-\frac{V_{0}}{%
r^{2}}\chi _{1}+\left( \chi _{1}^{2}+\gamma \chi _{2}^{2}\right) \frac{\chi
_{1}}{r^{2}},  \notag \\
&&  \label{stat} \\
\mu _{2}\chi _{2} &=&-\frac{1}{2}\chi _{2}^{\prime \prime }-\frac{V_{0}}{%
r^{2}}\chi _{2}+\left( \chi _{2}^{2}+\gamma \chi _{1}^{2}\right) \frac{\chi
_{2}}{r^{2}}.  \notag
\end{eqnarray}%
In terms of functions $\chi _{n}(r)$, the norms of the wave functions are%
\begin{equation}
N_{n}\equiv \int \left\vert \phi _{n}(\mathbf{r})\right\vert d\mathbf{r}%
=4\pi \int_{0}^{\infty }\left[ \chi _{n}(r)\right] ^{2}dr,  \label{N}
\end{equation}%
and the average squared radial size of the trapped mode is%
\begin{equation}
\left\langle r_{n}^{2}\right\rangle =\frac{\int_{0}^{\infty }\left[ \chi
_{n}(r)\right] ^{2}r^{2}dr}{\int_{0}^{\infty }\left[ \chi _{n}(r)\right]
^{2}dr}.  \label{<>}
\end{equation}

An expansion of solutions to Eqs.~(\ref{stat}) at $r\rightarrow 0$ is looked
for as \cite{we1}
\begin{equation}
\chi _{n}(r)=\chi _{n}^{(0)}\left[
1-c_{n}^{(1)}r^{s/2}-c_{n}^{(2)}r^{s/2+2}+\cdots
-d_{n}^{(1)}r^{2}-d_{n}^{(2)}r^{4}+\cdots \right] ,  \label{chi}
\end{equation}%
with a positive power, $s>0$ (here, $c_{1}\neq c_{2}$ is possible, but power
$s$ must be the same for $\chi _{1}$ and $\chi _{2}$). The terms $\sim
c_{n}^{(1,2,...)}$ and $d_{n}^{(1,2,...)}$ are produced, respectively, by
terms on the right- and left-hand sides of Eqs. (\ref{stat}). The
leading-order (most singular) terms, $\sim r^{-2}$, produced by the
substitution of ansatz (\ref{chi}) into Eqs. (\ref{stat}), lead to a system
of algebraic relations for coefficients $\chi _{n}^{(0)}$:
\begin{eqnarray}
\chi _{1}^{(0)}\left[ \left( \chi _{1}^{(0)}\right) ^{2}+\gamma \left( \chi
_{2}^{(0)}\right) ^{2}\right] &=&V_{0}\chi _{1}^{(0)},  \notag \\
&&  \label{chichi} \\
\chi _{2}^{(0)}\left[ \left( \chi _{2}^{(0)}\right) ^{2}+\gamma \left( \chi
_{1}^{(0)}\right) ^{2}\right] &=&V_{0}\chi _{2}^{(0)}.  \notag
\end{eqnarray}%
Obviously, Eqs. (\ref{chichi}) may have solutions of two types,
corresponding to miscible and immiscible states, respectively:%
\begin{eqnarray}
\chi _{1}^{(0)} &=&\chi _{2}^{(0)}\equiv \chi _{\mathrm{misc}}^{(0)}=\sqrt{%
V_{0}/\left( 1+\gamma \right) };  \label{mix} \\
\chi _{1}^{(0)} &\equiv &\chi _{\mathrm{immisc}}^{(0)}=\sqrt{V_{0}},~\chi
_{2}^{(0)}=0.  \label{demix}
\end{eqnarray}

The miscible and immiscible states, which are defined by relations (\ref{mix}%
) and (\ref{demix}), are relevant strictly for $\gamma <1$ and $\gamma >1$,
respectively. Our numerical analysis demonstrates that immiscible modes do
not exist at $\gamma <1$, while miscible ones are completely unstable at $%
\gamma >1$. Thus, unlike other settings featuring the
miscibility-immiscibility transitions \cite{Merhasin}-\cite{SO}, in the
present situation the transition point undergoes no shift from $\gamma =1$.

In the miscible state, the substitution of ansatz (\ref{chi}) into Eqs. (\ref%
{stat}) identifies two \emph{independent} next orders, past $r^{-2}$, in the
expansion at $r\rightarrow 0$, namely, $r^{s/2-2}$ and $r^{0}$ (constant).
These terms are produced, respectively, by the right- and left-hand sides of
Eqs. (\ref{stat}), as said above. The analysis of the former order ($%
r^{s/2-2}$) yields power $s$,
\begin{equation}
s=1+\sqrt{1+16V_{0}},  \label{s}
\end{equation}%
with $c_{1}=c_{2}$ in Eq. (\ref{chi}), while coefficient $c_{1}$ itself
remains indefinite, in terms of the asymptotic expansion at $r\rightarrow 0$%
. The consideration of the latter order ($r^{0}$) produces coefficients $%
d_{n}^{(1)}$ in Eq. (\ref{chi}):
\begin{equation}
d_{n}^{(1)}=\frac{\mu _{1}+\mu _{2}}{2\left( 1-2V_{0}\right) }-\frac{%
(-1)^{n}\left( \mu _{1}-\mu _{2}\right) }{2\left( 1-2V_{0}\frac{1-\gamma }{%
1+\gamma }\right) }.  \label{r^2}
\end{equation}%
Furthermore, exactly at the \textit{critical point}, $\left( V_{0}^{(\mathrm{%
cr})}\right) _{2}=1/2$, at which the first term in expression (\ref{r^2})
does not make sense, expansion (\ref{chi}) for the fully mixed state is
replaced by%
\begin{equation}
\chi _{n}(r)=\frac{1}{\sqrt{2\left( 1+\gamma \right) }}\left[ 1+\frac{\mu
_{1}+\mu _{2}}{4}r^{2}\ln \left( \frac{r_{0}}{r}\right) +(-1)^{n}\frac{%
1+\gamma }{4\gamma }\left( \mu _{1}-\mu _{2}\right) r^{2}\right] ,
\label{ln}
\end{equation}%
where $r_{0}$ is an arbitrary scale constant, in terms of the expansion at $%
r\rightarrow 0$.

\begin{figure}[tbp]
\begin{center}
\includegraphics[height=4.cm]{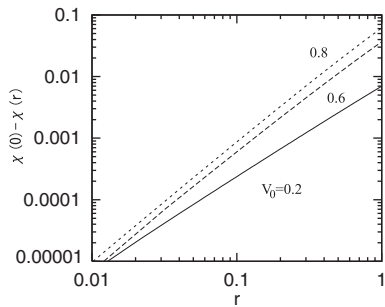}
\end{center}
\caption{Numerically found profiles of stationary wave functions for Eq.~(4) are shown, in the form of $\protect\chi (0)-\protect\chi (r)$, where $\protect\chi (0)=
\protect\sqrt{V_{0}}$ at $\gamma=0.9$  on the
double-log scale for several values of $V_{0}$, below and above the
weak-phase-transition-point, i.e, at $V_{0}<1/2$ and $V_{0}>1/2$,
respectively. All the profiles are obtained for the same norm, $N_{1}=4\protect\pi $.}
\label{fig1}
\end{figure}

These results demonstrate the \emph{breakup of analyticity} in the
dependence of the GS wave function on $V_{0}$ at $V_{0}=1/2$. Indeed, at $%
V_{0}<1/2$, i.e., $s/2<2$ [see Eq. (\ref{s})] the expansion of the wave
function at $r\rightarrow 0$ is dominated by terms $\sim r^{s/2}$, while at $%
V_{0}>1/2$ the power of the leading terms ($\sim r^{2}$) is fixed to be $2$,
i.e., the non-analyticity is manifest in the dependence of the power of the
leading term of the expansion on $V_{0}$. This conclusion is corroborated in
Fig. \ref{fig1} by shapes of functions $\chi (0)-\chi (r)\equiv \sqrt{V_{0}
}-\chi _{1}(r)$, found from numerical solutions of Eq.~(\ref{stat}) at $\gamma=0.9<1$ at different values of $V_{0}$. The power of the first correction to the constant term is given by
the slope of the plots shown on the double-logarithmic scale in Fig. \ref%
{fig1}. The slope is indeed equal to $2$ at $V_{0}\geq 1/2$, taking
smaller values at $V_{0}<1/2$. Note that at $V_{0}>1/2$ terms $\sim
r^{s/2}$ in expansion (\ref{chi}), although they are no longer
leading ones, are essential, as the free coefficient in front of
them, $c_{1}=c_{2}$, is necessary to adjust the wave function to the
condition of its exponential decay at $r\rightarrow \infty $, see
Eq. (\ref{exp}) below.

Non-analyticity of structural and correlation functions is a characteristic
feature of phase transitions in many-body settings, as shown in detail for
the Calogero-Sutherland model \cite{Grisha} and for many other systems \cite%
{nonanalytic}, including, incidentally, binary fluids \cite{binary-fluid}%
.Therefore, the analyticity breakup observed in the present model at $\left(
V_{0}^{(\mathrm{cr})}\right) _{2}=1/2$ suggests that the model undergoes a
phase transition at this point, although the GS exists equally well at $%
V_{0}<1/2$ and $V_{0}>1/2$. Recall that the onset of the quantum collapse in
the linear version of the model occurs at the critical value $\left( V_{0}^{(%
\mathrm{cr})}\right) _{1}=1/8$ \cite{LL}, which is a quarter of $\left(
V_{0}^{(\mathrm{cr})}\right) _{2}$. In fact, this phase transition is a
\emph{weak} one, as the structural characteristic of the wave function which
features the loss of the analyticity is a subtle one. Accordingly, it is
shown below (see Fig. \ref{fig6}) that the transition can be seen in a
subtle change of the experimentally observable dependence of the average
radius of the ground state [Eq. (\ref{<>})] on the attraction strength, $%
V_{0}$.

In the general case of the imbalanced miscible gas, with $\mu _{1}\neq \mu
_{2}$, Eq. (\ref{r^2}) demonstrates that a singularity in the GS wave
function occurs at an additional critical point,
\begin{equation}
\left( V_{0}^{(\mathrm{cr})}\right) _{3}=\frac{1}{2}\frac{1+\gamma }{%
1-\gamma }>\left( V_{0}^{(\mathrm{cr})}\right) _{2}.  \label{cr}
\end{equation}%
Thus, in addition to the phase transition in the balanced or imbalanced
bosonic gas at $V_{0}=1/2$, it is possible to expect another transition in
the imbalanced binary gas at point (\ref{cr}). Further analysis of this
issue is beyond the scope of the present work.

For the immiscible state, with $\gamma >1$, the expansion of $\chi _{1}(r)$
at $r\rightarrow 0$ is built as per Eqs. (\ref{chi}), (\ref{s}), (\ref{r^2}%
), and (\ref{ln}), with $\mu _{2}$ substituted by $\mu _{1}$ in the two
latter equations. However, due to the immiscibility, the expansion is
completely different for wave function $\chi _{2}(r)$:%
\begin{equation}
\chi _{2}(r)=\chi _{2}^{(1)}r^{s_{2}},~s_{2}=1+\sqrt{1+8\left( \gamma
-1\right) V_{0}},  \label{chi2}
\end{equation}%
which is valid for all values of $V_{0}>0$, while $\chi _{2}^{(1)}$ remains
an indefinite constant, in terms of the expansion at $r\rightarrow 0$.

Finally, at $r\rightarrow \infty $ Eqs. (\ref{stat}) yield an exponential
asymptotic form of the solution, which is valid for trapped modes of any
type,
\begin{equation}
\chi _{n}(r)\approx \chi _{n}^{(\infty )}\left( 1-\frac{V_{0}}{\sqrt{-2\mu
_{n}}r}\right) \exp \left( -\sqrt{-2\mu _{n}}r\right) ,  \label{exp}
\end{equation}%
where constants $\chi _{n}^{(\infty )}$ are indefinite in terms of the
asymptotic expansion at $r\rightarrow \infty $. Obviously, bound states,
that we aim to analyze here, may exist only for negative values of both
chemical potentials, $\mu _{1,2}<0$.

\section{Numerical and additional analytical results for trapped binary modes%
}

\subsection{Miscible ground states}

\begin{figure}[tbp]
\begin{center}
\includegraphics[height=4.cm]{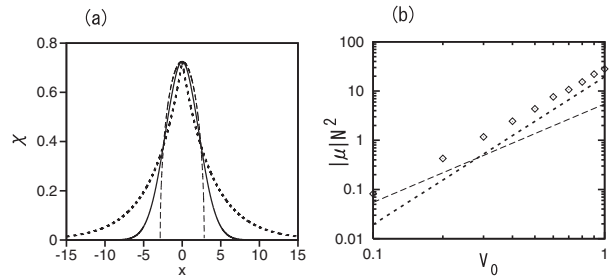}
\end{center}
\caption{(a) The numerically found profile of stationary wave functions $%
\protect\chi _{1}(r)=\protect\chi _{2}(r)$ at $V_{0}=1$ and $\protect\gamma %
=0.9$, with $N_{1}=N_{2}=4\protect\pi $, and its comparison with the
exponential analytical approximation given by Eq.~(\protect\ref{old}) and
the TF approximation Eq.~(\protect\ref{new}) (the short- and long-dashed
lines, respectively). (b) The chain of rhombuses depicts the numerically
found relation between $|\protect\mu |N^{2}$ and $V_{0}$ at $\protect\gamma %
=0.9$. The short- and long-dashed lines represent the approximations
provided by Eqs. (\protect\ref{oldN}) and (\protect\ref{e2}), respectively.}
\label{fig2}
\end{figure}

Figure \ref{fig2}(a) shows a typical example of the stationary profile for
the miscible GSs produced by a numerical solution of Eqs.~(\ref{stat}) at $
V_{0}=1$ for $\gamma =0.9$ and equal values of the norms of the two
components, $N_{1}=N_{2}=4\pi $, see Eq. (\ref{N}). The full density, $%
\left\vert \phi (r)\right\vert ^{2}=r^{-2}\chi ^{2}(r)$ [see Eq. (\ref{phi}%
)] monotonously diverges at $r\rightarrow 0$ (keeping the total norm
convergent).

The simplest global analytical approximation for the wave function of the GS
was proposed in Ref. \cite{we1} as an interpolation between the asymptotic
expansions (\ref{chi}), ignoring the corrections $\sim r^{s/2}$ and $r^{2}$,
and the leading term in expansion (\ref{exp}):
\begin{equation}
\chi _{n}(r)\approx \chi _{\mathrm{misc}}^{(0)}e^{-\sqrt{-2\mu _{n}}r}.
\label{old}
\end{equation}%
The substitution of this interpolation into Eqs. (\ref{N}) and (\ref{<>}),
along with expression (\ref{mix}), yields an approximate relations between
the chemical potentials and squared average radius of two components, and
their norms:
\begin{eqnarray}
\mu _{n} &=&-2\left[ \frac{\pi V_{0}}{(1+\gamma )N_{n}}\right] ^{2},
\label{oldN} \\
\left\langle r_{n}^{2}\right\rangle &=&\left[ \frac{(1+\gamma )N_{n}}{2\pi
V_{0}}\right] ^{2}.  \label{oldr^2}
\end{eqnarray}%
Comparison of Eq. (\ref{oldN}) with numerical results is shown in Fig.~\ref%
{fig2}(b) by the dashed line. This approximation is accurate for
sufficiently small $V_{0}$, but becomes inaccurate for large $V_{0}$.

For large $V_{0}$, the Thomas-Fermi (TF) approximation can be applied to the
\emph{balanced mixture}, with $N_{1}=N_{2}\equiv N$. In this approximation,
the derivative terms are neglected in Eq. (\ref{stat}), which yields (for $
\chi _{1}=\chi _{2}\equiv \chi $)
\begin{equation}
\chi _{\mathrm{TF}}(r)=\left\{
\begin{array}{c}
\sqrt{\left( V_{0}+\mu r^{2}\right) /\left( 1+\gamma \right) }\,,~~\mathrm{at%
}~~r<R_{0}\equiv \sqrt{V_{0}/(-\mu )}, \\
0,~~\mathrm{at}~~r>R_{0}~.%
\end{array}%
\right.  \label{new}
\end{equation}%
The substitution of approximation (\ref{new}) into Eqs. (\ref{N}) and (\ref%
{<>}) yields the corresponding $\mu _{\mathrm{TF}}(N)$ and $\left\langle r_{%
\mathrm{TF}}^{2}\right\rangle (N)$ relations,
\begin{eqnarray}
\mu _{\mathrm{TF}} &=&-\frac{64\pi ^{2}V_{0}^{3}}{9(1+\gamma )^{2}N^{2}},
\label{e2} \\
\left\langle r_{\mathrm{TF}}^{2}\right\rangle &=&\frac{5}{\pi ^{3}}\left[
\frac{3\left( 1+\gamma \right) N}{16V_{0}}\right] ^{2}\equiv \frac{5}{4\pi }%
R_{0}^{2},  \label{e3}
\end{eqnarray}%
where $N$ is the norm of one component [recall $R_{0}$ is the TF cutoff
radius defined in Eq. (\ref{new}); naturally, the average radius is smaller
than the cutoff value]. Analytical approximations (\ref{old}) and (\ref{new}%
) (shown by the short- and long-dashed lines, respectively) are compared
with the numerically found profile of the GS in Fig. \ref{fig2}(b),
demonstrating that the TF approximation provides for a good description of
the core part of the GS.

An obvious corollary of Eqs. (\ref{stat}) and (\ref{N}) are exact scaling
relations between $\mu $, $\left\langle r^{2}\right\rangle $ and $N$:
\begin{eqnarray}
\mu &\propto &-1/N^{2},  \label{scaling} \\
\left\langle r^{2}\right\rangle &\propto &N^{2},  \label{scaling2}
\end{eqnarray}%
for fixed $V_{0}$ and $\gamma $, cf. particular examples given by
Eqs. Note that scaling (\ref{scaling}) meets the
``anti-Vakhitov-Kolokolov" criterion, which plays the role of the
condition necessary for the stability of localized modes supported
by repulsive
nonlinearities \cite{we3}. Thus, product $|\mu |N^{2}$ does not depend on $N$%
, but does depend on the pull-to-the-center constant, $V_{0}$. This
numerically found dependence is shown by a chain of rhombuses in Fig. \ref%
{fig1}(b) on the double-logarithmic scale. It is seen that $|\mu |N^{2}$
increases with $V_{0}$ monotonously. In the same plot, the short- and
long-dashed lines show analytical approximations for the same dependence
given by Eq. (\ref{oldN}) and (\ref{e2}), respectively. It can be concluded
that, quite naturally, the TF approximation works better for larger $V_{0}$,
while the interpolation (\ref{old}) is more accurate for smaller $V_{0}$.

\begin{figure}[tbp]
\begin{center}
\includegraphics[height=4.cm]{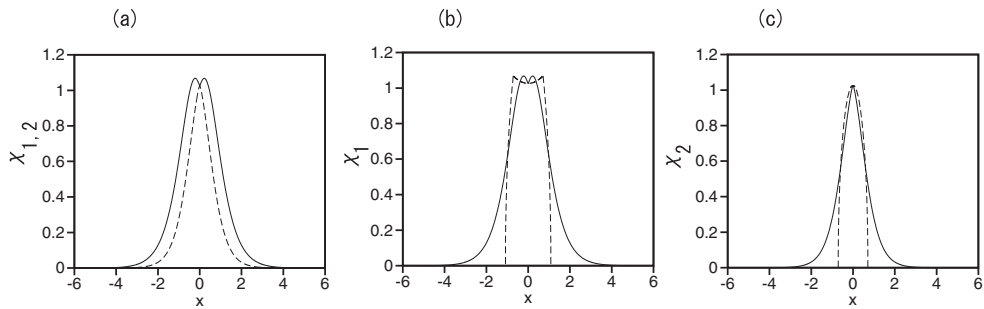}
\end{center}
\caption{(a) $\protect\chi _{1}$ (continuous) and $\protect\chi _{2}$
(dashed) components of the imbalanced miscible ground state at $V_{0}=2$ and
$\protect\gamma =0.9$, with $N_{1}=4\protect\pi $ and $N_{2}=2\protect\pi $.
(b) and (c): Comparison of the numerical result (continuous lines) with the
two-layer TF approximation (dashed lines) for $\protect\chi _{1}(r)$ and $%
\protect\chi _{2}(r)$. }
\label{fig3}
\end{figure}

Examples of numerically generated profiles of imbalanced miscible GSs is
shown in Fig.~\ref{fig3}(a) at $V_{0}=2$ and $\gamma =0.9$ for $N_{1}=4\pi $
and $N_{2}=2\pi $. Imbalanced miscible states with $\mu _{1}\neq \mu _{2}$
and $N_{1}\neq N_{2}$ are still characterized by equal values of $\chi
_{1,2}(r=0)$, in agreement with Eq. (\ref{mix}), but different constants $%
d_{1,2}^{(1)}$ in expansion (\ref{chi}), see Eq. (\ref{r^2}).

In the case of the strong pull to the center, $V_{0}\gg 1$, the TF
approximation, which has yielded Eqs. (\ref{new}) and (\ref{e2}) for the
balanced miscible GS, neglecting the derivatives in Eq. (\ref{stat}), can be
generalizes for imbalanced states, with $\mu _{1}\neq \mu _{2}$. For the
sake of the definiteness, we set $\mu _{2}\leq \mu _{1}$ (i.e., $\left\vert
\mu _{1}\right\vert \leq \left\vert \mu _{2}\right\vert $, as the chemical
potentials of the bound states are negative).

The TF approximation for imbalanced configuration is constructed in a
two-layer form, technically similar to that applied to the so-called
symbiotic gap solitons in Ref. \cite{Thawatchai}. In the \textit{inner layer}%
,
\begin{equation}
r^{2}<r_{0}^{2}\equiv \frac{1-\gamma }{\gamma \mu _{1}-\mu _{2}}V_{0},
\label{r0}
\end{equation}%
both wave functions are different from zero:%
\begin{equation}
\chi _{n}^{\mathrm{(inner)}}(r)=\sqrt{\frac{V_{0}}{1+\gamma }-\frac{\gamma
\mu _{3-n}-\mu _{n}}{1-\gamma ^{2}}r^{2}}.  \label{TF}
\end{equation}%
In the \textit{outer layer}, only one component is present, in the framework
of the TF approximation: $\chi _{2}\equiv 0$,%
\begin{equation}
\chi _{1}^{\mathrm{(outer)}}(r)=\left\{
\begin{array}{c}
\sqrt{V_{0}+\mu _{1}r^{2}},~\mathrm{at}~~r_{0}^{2}\leq r^{2}\leq
R_{0}^{2}\equiv -V_{0}/\mu _{1}, \\
0,~\mathrm{at}~~r^{2}\geq R_{0}^{2}~.%
\end{array}%
\right.  \label{out}
\end{equation}%
Note that both components of the TF solution, given by Eqs. (\ref{r0})-(\ref%
{out}), are continuous at $r=r_{0}$ and $r=R_{0}$. It is also worthy to note
that $\chi _{1}^{\mathrm{(inner)}}(r)$ is a decreasing or increasing
function of $r$ in the cases of $\left\vert \mu _{1}\right\vert <\left\vert
\mu _{2}\right\vert <|\mu _{1}|/\gamma $ and $\left\vert \mu _{2}\right\vert
>|\mu _{1}|/\gamma $, respectively. The two-layer TF approximation for a
typical imbalanced GS is compared to its numerical counterpart in Figs. \ref%
{fig3}(b,c).

The corresponding approximation for the dependence between the chemical
potentials and norms of the second component is produced by the
straightforward integration of expression (\ref{TF}), cf. Eq. (\ref{N}):%
\begin{equation}
N_{2}^{\mathrm{(TF)}}(\mu _{1},\mu _{2})=4\pi \int_{0}^{r_{0}}\left[ \chi
_{2}^{\mathrm{(inner)}}(r)\right] ^{2}dr=\frac{8\pi }{3}\frac{\sqrt{1-\gamma
}}{1+\gamma }V_{0}^{3/2}\left( \gamma \mu _{1}-\mu _{2}\right) ^{-1/2}.
\label{NTF}
\end{equation}%
The corresponding expression for the norm of the first component is very
cumbersome, therefore we give it here only for the limit case of $N_{1}\gg
N_{2}$, i.e., $\left\vert \mu _{2}\right\vert \gg \left\vert \mu
_{1}\right\vert $:%
\begin{equation}
N_{1}^{\mathrm{(TF)}}\approx \frac{8\pi }{3}V_{0}^{3/2}\left( -\mu
_{1}\right) ^{-1/2}.  \label{NTF1}
\end{equation}%
Note that dependences (\ref{NTF}) and (\ref{NTF1}) obey the generic scaling
law (\ref{scaling}). Expressions for the TF radii of the two imbalanced
components can be derived too, cf. Eq. (\ref{e3}), but they are very
cumbersome.

\subsection{Immiscible ground states}

\begin{figure}[tbp]
\begin{center}
\includegraphics[height=4.cm]{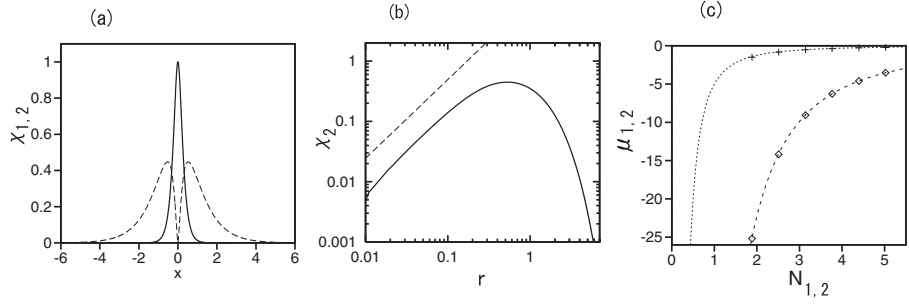}
\end{center}
\caption{(a) An example of numerically found profiles $\protect\chi _{1}(r)$
and $\protect\chi _{2}(r)$ (the solid and dashed lines) of the immiscible
ground state for $V_{0}=1,\protect\gamma =1.2$ and equal norms of the two
components, $N_{1}=N_{2}=0.8\protect\pi $, while $\protect\mu _{1}=-14.2$
and $\protect\mu _{2}=-0.84$. (b) The double-logarithmic plot of the same
component $\protect\chi _{2}(r)$. The dashed straight line shows, for the
sake of the comparison, the analytical prediction provided by Eq. (\protect
\ref{chi2}). (c) Chains of symbols depict relations $\protect\mu %
_{1,2}(N_{1,2})$ for the immiscible ground states at $V_{0}=1,\protect\gamma %
=1.2$. Dashed lines are empiric fits to scaling (\protect\ref{scaling}), $%
\protect\mu _{1}=-89/N^{2}$ and $\protect\mu _{2}=-5/N^{2}$.}
\label{fig4}
\end{figure}

As said above, in the case of $\gamma >1$ relevant states are immiscible
ones, which are approximated, at small $r$, by Eqs. (\ref{chi}), (\ref{s}), (%
\ref{r^2}), and (\ref{ln}) for $\chi _{1}(r)$, with $\mu _{2}$ substituted
by $\mu _{1}$ in the two latter equations, and by Eq. (\ref{chi2}) for $\chi
_{2}(r)$. A generic example of the immiscible GS found at $V_{0}=1,\gamma
=1.2$ and $N_{1}=N_{2}=0.8\pi $ is displayed in Fig. \ref{fig4}(a), and Fig. 
\ref{fig4}(b) shows the double-logarithmic plot of $\chi _{2}(r)$ at $%
V_{0}=1 $, comparing it to the analytical prediction given by Eq. (\ref{chi2}%
). Further, Fig. \ref{fig4}(c) displays relations between chemical
potentials and norms of both components of the immiscible GSs. The dashed
curves in the latter figure verify the validity of scaling relation (\ref%
{scaling}) in the present case.

\begin{figure}[tbp]
\begin{center}
\includegraphics[height=4.cm]{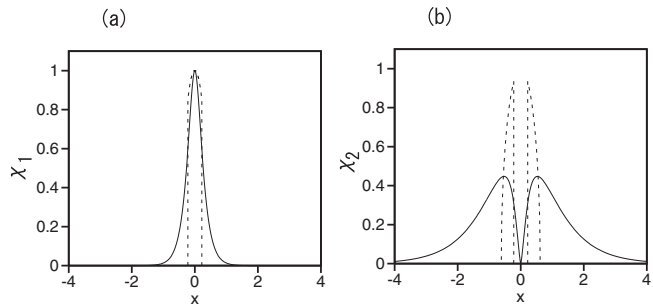}
\end{center}
\caption{(a) and (b): The comparison of the numerically found profiles for
components $\protect\chi _{1}(r)$ and $\protect\chi _{2}(r)$ of the
immiscible ground state, shown in Fig.~\protect\ref{fig4}(a) (solid lines),
with the respective TF approximations, given by Eqs. (\protect\ref{ImmTFin})
and (\protect\ref{ImmTFout}), respectively (continuous lines).}
\label{fig5}
\end{figure}

The TF approximation produces a two-layer solution for the immiscible state.
In the inner layer,
\begin{equation}
r^{2}<r_{0}^{2}=\frac{(\gamma -1)V_{0}}{\gamma \mu _{1}-\mu _{2}},  \notag
\end{equation}%
the approximation yields%
\begin{equation}
\chi _{1}(r)=\sqrt{V_{0}+\mu _{1}r^{2}},\;\chi _{2}(r)=0.  \label{ImmTFin}
\end{equation}%
In the outer layer, with $r_{0}^{2}<r^{2}<R_{0}^{2}=V_{0}/(-\mu _{2})$, the
result is
\begin{equation}
\chi _{1}(r)=0,\;\chi _{2}(r)=\sqrt{V_{0}+\mu _{2}r^{2}}.  \label{ImmTFout}
\end{equation}%
In these solutions, $\mu _{1}$ and $\mu _{2}$ are related to the norms.
\begin{equation}
N_{1}^{(\mathrm{TF})}(\mu _{1},\mu _{2})=4\pi \int_{0}^{r_{0}}\chi
_{1}(r)^{2}dr=4\pi r_{0}\frac{\mu _{2}-(2/3)\gamma \mu _{1}-\mu _{1}/3}{\mu
_{2}-\gamma \mu _{1}},  \notag
\end{equation}%
\begin{gather}
N_{2}^{(\mathrm{TF})}(\mu _{1},\mu _{2})=4\pi \int_{r_{0}}^{R_{0}}\chi
_{2}^{2}(r)dr  \notag \\
=4\pi \left[ \frac{2}{3}\frac{V_{0}^{3/2}}{\sqrt{-\mu _{2}}}-\sqrt{\gamma -1}%
\left( \frac{V_{0}}{(\mu _{2}-\gamma \mu _{1}}\right) ^{3/2}\left( \frac{2}{3%
}\mu _{2}+\frac{1}{3}\gamma \mu _{2}-\gamma \mu _{1}\right) \right] .  \notag
\end{gather}%
Figure \ref{fig5} compares this TF approximation with numerical results.

\subsection{Characterization of the weak phase transition}

Finally, it is relevant to discuss how the weak phase transition, predicted
above at $V_{0}=\widetilde{V}_{0}^{(\mathrm{cr})}\equiv 1/2$, may manifest
itself in terms of experimentally measurable quantities. To this end, it is
relevant to consider dependences of the GS's characteristics on $V_{0}$ at a
fixed value of the norm, $N$, which is sufficient to do for the
single-component model (this problem was not considered previously in Ref.
\cite{we2}). In fact, scaling relations (\ref{scaling}) and (\ref{scaling2})
demonstrate that these dependences have the same form at all constant values
of $N$.

In the experiment, the variation of $V_{0}$ may be possible in the gas
composed of excited atoms: because the dipole moment of the atom in the
Rydberg state scales with the principal quantum number, $n$, as $n^{2}$, the
change $n\rightarrow n+1$ will give rise, according to Eq. (\ref{V0}), to a
relative variation of the attraction constant, $\left\vert \Delta
V_{0}\right\vert /V_{0}\approx 2/n$, which is small for sufficiently large $n
$.

\begin{figure}[tbp]
\begin{center}
\includegraphics[height=4.cm]{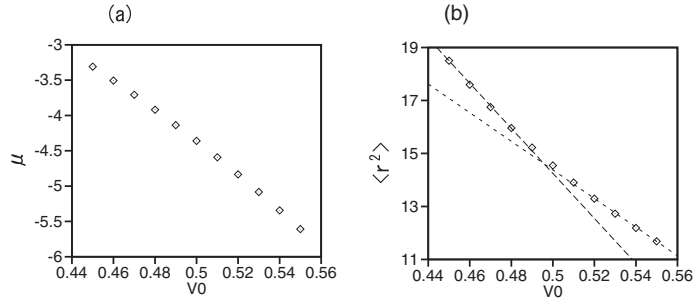}
\end{center}
\caption{(a) The dependence of the chemical potential on the attraction
strength, $V_{0}$, for a fixed norm, $N=4\protect\pi $, in the miscible state at $\gamma=0.9$. (b) The dependence of
the average squared radius of the ground state, calculated as per Eq. (\protect\ref{<>}), on $V_{0}$, in the same case. Straight tangent
lines illustrate the change in the slope of the dependence around
the weak-phase-transition point, $V_{0}=1/2$.} \label{fig6}
\end{figure}

First, Fig. \ref{fig6}(a) shows that the dependence of the chemical
potential on $V_{0}$ does not show any visible peculiarity at $V_{0}=1/2$
(however, the chemical potential is not an observable quantity). On the
other hand, the plot for $\left\langle r^{2}\right\rangle $ versus $V_{0}$,
displayed in Fig. \ref{fig6}(b), shows a subtle but observable feature:
the change of the slope of the dependence from $-85$ to $-54$ (in the scaled
units adopted in this work) in a small vicinity of the the phase-transition
point, as illustrated by means of the two tangent lines. Thus, the
experimental observation of the size of the ground state may reveal the
occurrence of the phase transition.

\section{The binary system with the single component pulled to the center}

Next, we consider the binary system with the attractive potential $%
-V_{0}/r^{2}$ acting only on the first species. In this case, we adopt the
combination of the intra-species self-repulsion and attraction between the
species, i.e., $\gamma <0$ in the accordingly modified system of Eqs. (\ref%
{GP}) and (\ref{stat}):
\begin{eqnarray}
i\frac{\partial \phi _{1}}{\partial t} &=&-\frac{1}{2}\nabla ^{2}\phi
_{1}+(|\phi |^{2}+\gamma |\phi _{2}|^{2})\phi _{1}-\frac{V_{0}}{r^{2}}\phi
_{1},  \notag \\
&&  \label{GP1} \\
i\frac{\partial \phi _{2}}{\partial t} &=&-\frac{1}{2}\nabla ^{2}\phi
_{2}+(\gamma |\phi _{1}|^{2}+|\phi _{2}|^{2})\phi _{2},  \notag
\end{eqnarray}%
\begin{eqnarray}
\mu _{1}\chi _{1} &=&-\frac{1}{2}\chi _{1}^{\prime \prime }-\frac{V_{0}}{%
r^{2}}\chi _{1}+(\chi _{1}^{2}+\gamma \chi _{2}^{2})\frac{\chi _{1}}{r^{2}},
\notag \\
&&  \label{stat1} \\
\mu _{2}\chi _{2} &=&-\frac{1}{2}\chi _{2}^{\prime \prime }+(\chi
_{2}^{2}+\gamma \chi _{1}^{2})\frac{\chi _{2}}{r^{2}}.  \notag
\end{eqnarray}

Solutions to Eqs. (\ref{stat1}) can be again sought for in the form of
expansion (\ref{chi}) at $r\rightarrow 0$, which yields, instead of Eqs. (%
\ref{chichi}), a system of algebraic relations
\begin{eqnarray}
\chi _{1}^{(0)}\left[ \left( \chi _{1}^{(0)}\right) ^{2}+\gamma \left( \chi
_{2}^{(0)}\right) ^{2}\right]  &=&V_{0}\chi _{1}^{(0)},  \notag \\
&&  \label{chichi1} \\
\chi _{2}^{(0)}\left[ \left( \chi _{2}^{(0)}\right) ^{2}+\gamma \left( \chi
_{1}^{(0)}\right) ^{2}\right]  &=&0.  \notag
\end{eqnarray}%
It is easy to see that, in the present case with $\gamma <0$, a nontrivial
solution to Eqs. (\ref{chichi1}) exists for $|\gamma |~<1$:
\begin{equation}
\chi _{1}^{(0)}=\sqrt{V_{0}/(1-\gamma ^{2})},~\chi _{2}^{(0)}=\sqrt{-\gamma }%
\chi _{10}.  \label{chi12}
\end{equation}

Further, the interpolation ansatz, following the pattern of Eq. (\ref{old}),%
\begin{equation}
\chi _{n}(x)=\chi _{n}^{(0)}\exp \left( -\sqrt{-2\mu _{n}}r\right) ,
\label{old1}
\end{equation}%
with $\chi _{n}^{(0)}$ given by Eq. (\ref{chi12}), predicts the following
relations between the chemical potentials and norms of the two components:
\begin{equation}
\mu _{1}=-\frac{1}{2}\left( \frac{2\pi V_{0}N_{1}}{1-\gamma ^{2}}\right)
^{2},\ \mu _{2}=-\frac{1}{2}\left( \frac{2\pi \gamma V_{0}N_{2}}{1-\gamma
^{2}}\right) ^{2}.  \label{mu12}
\end{equation}

\begin{figure}[tbp]
\begin{center}
\includegraphics[height=4.cm]{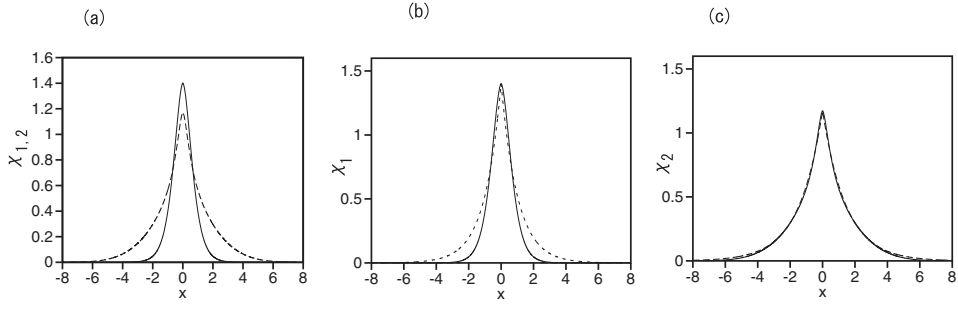}
\end{center}
\caption{(a) Numerically found profiles of stationary wave functions $%
\protect\chi _{1}(r)$ and $\protect\chi _{2}(r)$ (the taller and lower ones,
respectively) in the model based on Eqs. (\protect\ref{GP1}) and (\protect
\ref{stat1}), for $V_{0}=1,\protect\gamma =-0.7$. $\protect\mu _{1}=-2.59$
and $\protect\mu _{2}=-0.18$. (b) and (c): The comparison of $\protect\chi %
_{1}(r)$ and $\protect\chi _{2}(r)$ with the analytical approximations
provided by Eq. (\protect\ref{old1}).}
\label{fig7}
\end{figure}

Figure \ref{fig7}(a) displays a typical example of the numerically found
profiles of $\chi _{1}(r)$ and $\chi _{2}(r)$, for $V_{0}=1,\gamma
=-0.7,N=4\pi $. Further, Figs. \ref{fig7}(b) and (c) compare the numerical
profiles of $\chi _{1}(r)$ and $\chi _{2}(r)$ with the analytical
approximations based on \textit{ans\"{a}tze} (\ref{new}), (\ref{chi12}) and (%
\ref{old1}).

\begin{figure}[tbp]
\begin{center}
\includegraphics[height=4.cm]{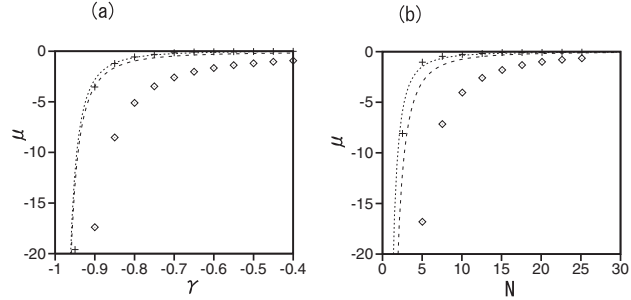}
\end{center}
\caption{(a) The chemical potentials of the two components, $\protect\mu _{1}
$ ($\Diamond $) and $\protect\mu _{2}$ ($+$), in the model based on Eqs. (%
\protect\ref{GP1}) and (\protect\ref{stat1}), versus $\protect\gamma $ for $%
N_{1}=N_{2}=4\protect\pi $ and $V_{0}=1$. (b) The same chemical potentials
versus $N_{1}=N_{2}\equiv N$ for $\protect\gamma =-0.7$ and $V_{0}=1$. The
dashed and dotted curves show the analytical prediction (\protect\ref{mu12}%
). }
\label{fig8}
\end{figure}

The family of the GSs in the present version of the binary system is
characterized by dependences of the chemical potentials of the two
components on coupling constant $\gamma $, see Fig. \ref{fig8}(a), and on $%
N_{1}=N_{2}\equiv N$ in Fig. \ref{fig8}(b). In both panels, the dashed and
dotted lines depict the analytical prediction for $\mu _{1}$ and $\mu _{2}$
produced by Eq. (\ref{mu12}). Thus, it is concluded that the analytical
approximation provide for quite an accurate description of the $\chi _{2}$
component (the one which is not pulled to the center), while the
approximation for $\chi _{1}$ is less accurate.

Finally, the two-layer TF\ approximation can be applied to the present
system too, under conditions $\gamma ^{2}<1$ and $\mu _{2}<\mu _{1}<0$. In
particular, the solution in the inner layer,
\begin{equation}
r^{2}\leq r_{0}^{2}\equiv -\frac{|\gamma |}{\mu _{2}+|\gamma |\mu _{1}}V_{0},
\notag
\end{equation}%
cf. Eq. (\ref{r0}), takes the following form, cf. Eq. (\ref{TF}):%
\begin{eqnarray}
\chi _{1}^{\mathrm{(inner)}}(r) &=&\sqrt{\frac{V_{0}+\left( \mu _{1}+|\gamma
|\mu _{2}\right) r^{2}}{1-\gamma ^{2}}},  \notag \\
\chi _{2}^{\mathrm{(inner)}}(r) &=&\sqrt{\frac{|\gamma |V_{0}+\left( \mu
_{2}+|\gamma |\mu _{1}\right) r^{2}}{1-\gamma ^{2}}},  \notag
\end{eqnarray}%
the corresponding dependence of the norm on the chemical potentials being%
\begin{equation}
N_{2}^{\mathrm{(TF)}}(\mu _{1},\mu _{2})=\frac{8\pi }{3}\left( |\gamma
|V_{0}\right) ^{2/3}\left[ -\left( \mu _{2}+|\gamma |\mu _{1}\right) \right]
^{-1/2},  \notag
\end{equation}%
cf. Eq. (\ref{NTF}). In the outer layer, the TF solution takes the same from
as in Eq. (\ref{out}).

\section{Conclusion}

The aim of this work is to extend the recent analysis of the suppression of
the quantum collapse, induced by action of the attractive potential $%
-V_{0}r^{-2}$ in the 3D space, by the repulsive nonlinearity in the bosonic
gas, to the binary miscible and immiscible gases. We have concluded that the
GS (ground state) which the respective linear Schr\"{o}dinger equation fails
to create, emerges in the nonlinear system. The GSs were found in the
numerical form, and approximated by means of several analytical methods,
such as the TF (Thomas-Fermi) approximation, which is relevant for large $%
V_{0}$. An essential finding is that both balanced and imbalanced miscible
gases feature the weak phase transition in the GS, in the form of the
analyticity-breaking change in the structure of the wave functions at small $%
r$. The binary system with the attraction between the species (rather than
repulsion) has been considered too, in the case when the attractive central
potential acts on a single component.

\section*{Acknowledgment}

We appreciate a valuable discussions with G. E. Astrakharchik and J.
Schmiedmayer.

\end{document}